\documentclass[twocolumn,showpacs,preprintnumbers,amsmath,amssymb]{revtex4}

\usepackage{graphicx}
\usepackage{dcolumn}
\usepackage{bm}

\newcommand{\bq}{\begin{equation}}
\newcommand{\eq}{\end{equation}}
\newcommand{\bqa}{\begin{eqnarray}}
\newcommand{\eqa}{\end{eqnarray}}
\newcommand{\nn}{\nonumber \\}

\def\be     {\begin{equation}}
\def\ee     {\end{equation}}
\def\bea        {\begin{eqnarray}}
\def\eea        {\end{eqnarray}}
\def\bnn    {\begin{eqnarray*}}
\def\enn    {\end{eqnarray*}}

\begin{document}

\title{Deconfined Quantum Criticality at the Quantum Phase Transition from Antiferromagnetism to Algebraic Spin Liquid}
\author{Ki-Seok Kim}
\affiliation{ School of Physics, Korea Institute for Advanced
Study, Seoul 130-012, Korea }
\date{\today}

\begin{abstract}
We investigate the quantum phase transition from
antiferromagnetism ($AF$) to algebraic spin liquid ($ASL$). {\it
We propose that spin $1/2$ fermionic spinons in the $ASL$
fractionalize into spin $1/2$ bosonic spinons and spinless
fermions at the quantum critical point ($QCP$) between the $AF$
and the $ASL$}. Condensation of the bosonic spinons leads to the
$AF$, where the condensed bosonic spinons are confined with the
spinless fermions to form the fermionic spinons. These fermionic
spinons are also confined to make antiferromagnons as elementary
excitations in the $AF$. {\it Approaching the $QCP$ from the $AF$,
spin $1$ critical antiferromagnetic fluctuations are expected to
break up into spin $1/2$ critical bosonic spinons. Then, these
bosonic spinons hybridize with spin $1/2$ fermionic spinons,
making spinless fermions}. As a result the fermionic spinons decay
into the bosonic spinons and the spinless fermions. But, the
spinless fermions are confined and thus, only the bosonic spinons
emerge at the $QCP$. This coincides with the recent studies of
{\it deconfined quantum
criticality}\cite{Laughlin_deconfinement,Senthil_deconfinement,Kim1,Ichinose_deconfinement}.
When the bosonic spinons are gapped, the $ASL$ is realized. The
bosonic spinons are confined with the spinless fermions to form
the fermionic spinons. These fermionic spinons are deconfined to
describe the $ASL$.
\end{abstract}


\maketitle

\section{Introduction}

Nature of quantum criticality is one of the central interests in
modern condensed matter physics. Especially, {\it deconfined
quantum criticality} has been proposed in various strongly
correlated electron systems such as low dimensional quantum
antiferromagnetism on square
lattices\cite{Laughlin_deconfinement,Senthil_deconfinement,
Kim1,Ichinose_deconfinement,Kleinert1,Review,Hermele_QED3,Kleinert2,Kim2}
and geometrically frustrated lattices\cite{Kim3,SL_Review,SL1,SL2}
and heavy fermion
liquids\cite{Senthil_Kondo,Coleman_Kondo,Pepin_Kondo,Kim_Kondo1,Kim_Kondo2}.
In the present paper we focus our attention on the quantum
antiferromagnetism in two dimensional square lattices. Starting
from the antiferromagnetic Heisenberg model, one can derive the
O(3) nonlinear $\sigma$ model ($NL\sigma{M}$) as an effective
field theory. Utilizing the $CP^1$ representation of the
$NL\sigma{M}$, one can show that although the appropriate {\it
off-critical} elementary degrees of freedom are given by either
spin $1$ excitons (gapped paramagnons) in the paramagnetism or
spin $1$ antiferromagnons in the antiferromagnetism, {\it at the
quantum critical point} ($QCP$) such excitations {\it break up}
into more elementary spin $1/2$ excitations called {\it
spinons}\cite{Laughlin_deconfinement,Senthil_deconfinement,Kim1,
Ichinose_deconfinement}. This is the precise meaning of the
deconfined quantum criticality in the context of quantum
antiferromagnetism.

In the present paper we investigate the quantum phase transition
from antiferromagnetism ($AF$) to algebraic spin liquid ($ASL$)
{\it based on the deconfined quantum criticality of the O(3)
$NL\sigma{M}$}\cite{Laughlin_deconfinement,
Senthil_deconfinement,Kim1,Ichinose_deconfinement}. According to
one possible scenario\cite{Review}, the pseudogap phase in high
$T_c$ cuprates is proposed to be the $ASL$ for spin degrees of
freedom. In this respect it is considerable to understand how the
$ASL$ emerges from the $AF$, i.e., the quantum phase transition
between the $AF$ and the $ASL$. The $ASL$ is the state described
by quantum electrodynamics in two space and one time dimensions
($QED_3$) in terms of massless Dirac spinons interacting via U(1)
gauge fluctuations\cite{Review}. On the other hand, low energy
physics of the $AF$ is well understood by the O(3) $NL\sigma{M}$
in terms of Neel order parameter fields\cite{Chakravarty}. One can
say that the $AF$ results from the $ASL$ via spontaneous chiral
symmetry breaking
($S\chi{S}B$)\cite{CSB1,CSB2,DonKim_QED3,Herbut,Dmitri}. The
central question in this paper is the nature of the $QCP$, where
the $S\chi{S}B$ occurs. Concretely speaking, we examine how the
Neel order parameter fields in the $NL\sigma{M}$ can be smoothly
connected to the fermionic spinons in the $QED_3$ by studying the
$QCP$. We apply the deconfined quantum criticality of the O(3)
$NL\sigma{M}$ to the $QCP$ between the $AF$ and the $ASL$.
Approaching the $QCP$ from the $AF$, critical antiferromagnetic
spin fluctuations in the $NL\sigma{M}$ are conjectured to break up
into critical bosonic spinons, consistent with the previous
studies\cite{Laughlin_deconfinement,
Senthil_deconfinement,Kim1,Ichinose_deconfinement}. {\it Based on
this conjecture we propose that the critical bosonic spinons
hybridize with the fermionic spinons in the $QED_3$ via Kondo-like
couplings, screening out spin degrees of freedom of the fermionic
spinons and thus, making spinless fermions}. In other words, it
can be stated that the fermionic spinons in the $ASL$
fractionalize into the bosonic spinons and the spinless fermions
at the $QCP$. In this paper we construct a critical field theory
at the $QCP$ and discuss how the $AF$ and the $ASL$ recover from
the critical field theory.

In the language of high energy physics the present study examines
the chiral phase transition in $QED_3$. In this respect we would
like to point out an interesting previous study of the chiral
phase transition\cite{Appelquist}. In the study\cite{Appelquist}
the authors calculated a fermion-antifermion scattering amplitude
in order to find light scalar mesons (fermion-antifermion
composites) showing up as poles in the scattering amplitude.
Approaching the $QCP$ from the symmetric phase ($ASL$ in the
context of quantum antiferromagnetism), the authors could not find
the poles in the scattering amplitude. {\it This leads them to the
conclusion that there are no light mesons corresponding to order
parameter fluctuations}. Based on this result they proposed that
the chiral phase transition is different from usual second order
phase transitions described by Landau-Ginzburg-Wilson type
theories of order parameter fluctuations. Our present scenario of
the deconfined $QCP$ for the chiral phase transition seems to be
consistent with this previous study since the deconfined $QCP$ is
not described by fluctuations of fermion-antifermion composite
order parameters owing to spin fractionalization.

\section{Fermionic nonlinear $\sigma$ model as an effective field
theory for quantum antiferromagnetism}

We consider the following effective action called fermionic
nonlinear $\sigma$ model   \bqa && Z = \int{D{\vec
n}}{D\lambda}{D\psi_{\sigma}}{Da_{\mu}}
e^{-S_{ASL}-S_{M}-S_{NL\sigma{M}}} , \nn && S_{ASL} = \int{d^3x}
\Bigl[ \bar{\psi}_{\sigma}\gamma_{\mu}(\partial_{\mu} -
ia_{\mu})\psi_{\sigma} + \frac{1}{2e^2}|\partial\times{a}|^2
\Bigr] , \nn && S_{M} = \int{d^3x}
m_{\psi}\bar{\psi}_{\sigma}{\vec
\tau}_{\sigma\sigma'}\psi_{\sigma'}\cdot{\vec n} , \nn &&
S_{NL\sigma{M}} = \int{d^3x}
\Bigl[\frac{Nm_{\psi}}{4\pi}|\partial_{\mu}{\vec n}|^{2} -
i\lambda(|{\vec n}|^{2} - 1) \Bigr] . \eqa In $S_{ASL}$
$\psi_{\sigma}$ is a massless Dirac spinon with a flavor index
$\sigma = 1, ..., N$ associated with SU(N) spin symmetry.
$a_{\mu}$ is a {\it compact} U(1) gauge field mediating long range
interactions between Dirac spinons. $e$ is an {\it internal}
electric charge of the Dirac spinon. $S_{ASL}$ in Eq. (1) was
proposed to be an effective field theory for one possible quantum
paramagnetism ($ASL$) of SU(N) quantum antiferromagnets on two
dimensional square lattices\cite{Review}. However, the stability
of the $ASL$ has been suspected owing to instanton excitations of
compact U(1) gauge fields\cite{Herbut_confinement}. Condensation
of instantons (magnetic monopoles) is believed to cause
confinement of charged particles\cite{Polyakov,Fradkin,NaLee},
here Dirac spinons. Recently, Hermele et al. showed that the $ASL$
can be stable against instanton excitations\cite{Hermele_QED3}.
Ignoring the compactness of U(1) gauge fields $a_{\mu}$, one can
show that the $S_{ASL}$ has a nontrivial charged fixed point in
two space and one time dimensions [$(2+1)D$] in the limit of large
flavors\cite{Hermele_QED3,Kim2,Kleinert2}, identified with the
$ASL$. Hermele et al. showed that the charged critical point in
the case of noncompact U(1) gauge fields can be stable against
instanton excitations of compact U(1) gauge fields in the limit of
large flavors\cite{Hermele_QED3}. Condensation of instantons can
be forbidden at the stable charged fixed point owing to critical
fluctuations of Dirac fermions. The $S_{ASL}$ in Eq. (1) is a
critical field theory at the charged critical point, where
correlation functions exhibit power law behaviors with anomalous
critical exponents resulting from long range gauge
interactions\cite{Kim2,QED3_exponent}. This is the reason why the
state described by the $S_{ASL}$ is called the $ASL$. In appendix
A we briefly discuss how the effective $QED_3$, $S_{ASL}$ in Eq.
(1) can be derived from the antiferromagnetic Heisenberg model.

However, we should remember that the $ASL$ criticality holds only
for large flavors of critical Dirac spinons. If the flavor number
is not sufficiently large, the internal charge $e$ is not screened
out satisfactorily by critical Dirac fermions. Then, gauge
interactions can make bound states of Dirac fermions, resulting in
massive Dirac spinons. This is known to be spontaneous chiral
symmetry breaking
($S\chi{S}B$)\cite{CSB1,CSB2,DonKim_QED3,Herbut,Dmitri}. In the
case of physical SU(2) antiferromagnets it is not clear if the
$ASL$ criticality remains stable owing to the $S\chi{S}B$ causing
antiferromagnetism. It is believed that there exists the critical
flavor number $N_c$ associated with $S\chi{S}B$ in
$QED_3$\cite{CSB1,CSB2}. But, the precise value of the critical
number is far from consensus\cite{CSB2}. If the critical value is
larger than $2$, the $S\chi{S}B$ is expected to occur for the
physical $N = 2$ case, resulting in massive spinons. On the other
hand, in the case of $N_c < 2$ the $ASL$ criticality remains
stable against the $S\chi{S}B$. Experimentally, an
antiferromagnetic long range order is observed in the SU(2)
antiferromagnet. This leads us to consider the $S\chi{S}B$ in the
$ASL$. In Eq. (1) $S_{M}$ represents the contribution of a fermion
mass due to the $S\chi{S}B$. $m_{\psi}$ is a mass parameter
corresponding to staggered magnetization. ${\vec n}$ represents
fluctuations of Neel order parameter fields regarded as Goldstone
bosons in the $S\chi{S}B$. ${\vec \tau}$ is Pauli matrix acting on
the spin (flavor) space of Dirac spinons. The mass parameter
$m_{\psi}$ can be determined by a self-consistent gap equation,
given by $m_{\psi} \approx e^{2}exp[-2\pi/\sqrt{N_c/N - 1}]$ in
the $1/N$ approximation\cite{CSB1}. Varying the flavor number $N$
controls the mass parameter $m_{\psi}$. {\it In the present paper
we use the mass $m_{\psi}$ as a controlling parameter of the
quantum phase transition}. This is analogous to a quantum phase
transition in the $NL\sigma{M}$, where the quantum phase
transition is realized by varying the spin stiffness parameter
associated with the value of spin in the system\cite{Nagaosa}. For
completeness of this paper we briefly sketch the derivation of
dynamical mass generation in appendix B. There are additional
fermion bilinears connected with the Neel state by "chiral"
transformations because the $S_{ASL}$ in Eq. (1) has more
symmetries than those of the Heisenberg
model\cite{Herbut,Tanaka_sigma,Hermele_sigma}. These order
parameters are associated with valance bond
orders\cite{Herbut,Tanaka_sigma,Hermele_sigma}. But, in the
present paper we consider only the Neel order parameter in order
to take account of the $AF$ described by the O(3) $NL\sigma{M}$.

Contributions of high energy spinons in $S_{ASL} + S_{M}$ lead to
the $S_{NL\sigma{M}}$ in the gradient
expansion\cite{Kim_zeromode,Dmitri, Tanaka_sigma}. This effective
action is nothing but the O(3) $NL\sigma{M}$ describing a
collinear spin order. In the $S_{NL\sigma{M}}$ we introduced a
Lagrange multiplier field $\lambda$ to impose the rigid rotor
constraint $|{\vec n}|^2 = 1$. In appendix C we give a detailed
derivation of the $NL\sigma{M}$. The total effective action
$S_{ASL} + S_{M} + S_{NL\sigma{M}}$ called fermionic $NL\sigma{M}$
in Eq. (1) naturally describes the quantum $AF$ resulting from the
$ASL$ via $S\chi{S}B$ at half filling. Based on this fermionic
O(3) $NL\sigma{M}$ we investigate the $QCP$ of the $AF$ to $ASL$
transition.

In Eq. (1) we concentrate on the Kondo-like spin coupling term
$\vec{n}\cdot\bar{\psi}\vec{\tau}\psi$ between antiferromagnetic
spin fluctuations ${\vec n}$ and Dirac fermions $\psi_{\sigma}$.
This term shows that an antiferromagnetic excitation of spin $1$
can fractionalize into two fermionic spinons of spin $1/2$. But,
long range gauge interactions prohibit antiferromagnetic spin
fluctuations from decaying into spinons. Because massive Dirac
spinons in the $S\chi{S}B$ can generate only the Maxwell kinetic
energy for the gauge field $a_{\mu}$ via particle-hole
polarizations, they should be confined to form spin $1$
antiferromagnetic fluctuations owing to the effect of instantons.
Remember that the Maxwell gauge theory shows confinement of
charged matter fields in $(2+1)D$ owing to the condensation of
instantons\cite{Polyakov,Fradkin,NaLee}. As a result
spinon-antispinon composites appear to be identified with spin $1$
antiferromagnetic fluctuations, i.e., ${\vec n} =
\langle\bar{\psi}{\vec \tau}\psi\rangle$, where
$\langle...\rangle$ denotes a vacuum expectation value. A
resulting effective field theory for this antiferromagnet is
obtained to be \bqa && Z_{AF} = \int{D {\vec \pi}} e^{- S_{\pi}} ,
\nn && S_{\pi} = \int{d^3x} \frac{Nm_{\psi}}{4\pi}\Bigl(
|\partial_{\mu}{\vec \pi}|^{2} + \frac{({\vec
\pi}\cdot\partial_{\mu}{\vec \pi})^{2}}{1 - |{\vec \pi}|^{2}}
\Bigr) \nn && \approx \int{d^3x} \frac{Nm_{\psi}}{4\pi}\Bigl(
|\partial_{\mu}{\vec \pi}|^{2} + ({\vec
\pi}\cdot\partial_{\mu}{\vec \pi})^{2} \Bigr) , \eqa where the
Neel vector is represented as ${\vec n} = ({\vec \pi}, n^{3})$
with $n^{3} = \sqrt{1 - |{\vec \pi}|^2}$. In the last line we
obtained an effective field theory for small fluctuations of
${\vec \pi}$ fields around the Neel axis $n^{3}$ by using $n^{3}
\approx 1 - (1/2)|{\vec \pi}|^2$. The ${\vec \pi}$ fields are
nothing but aniferromagnons, considered to be spinon-antispinon
composites $\pi^{\pm} = \langle\bar{\psi}\tau^{\pm}\psi\rangle =
\pi_{1} \pm i\pi_{2}$ with the relativistic spectrum of $\omega =
k$ in the low energy limit. The last term in the last line
represents interactions between antiferromagnons. Low energy
physics in this conventional quantum antiferromagnet is well
described by the interacting antiferromagnons\cite{Chakravarty}.

\section{Quantum phase transition from antiferromagnetism to algebraic spin liquid}

\subsection{Effective Field Theory}

Reducing the mass parameter $m_{\psi}$ in Eq. (1) results in the
$QCP$ of the $AF$ to $ASL$ transition. It is important to realize
that although the Dirac spinons are confined to form
antiferromagnons in the $AF$, they should be deconfined to emerge
in the $ASL$. In order to investigate their confinement to
deconfinement transition, Dirac spinons should be explicitly
introduced in the effective action. Especially, it needs much care
to treat the $S_{M}$ in Eq. (1) representing spin
fractionalization. In the $AF$ the spinon mass can be considered
to be infinite owing to the confinement of spinons, allowing us to
ignore the process of spinon fractionalization. This permits us to
integrate over the spinon degrees of freedom completely and obtain
the effective $NL\sigma{M}$ Eq. (2) in terms of only the Neel
order parameter fields. On the other hand, near the $QCP$ this
full integration of spinons is difficult to be justified. Because
the mass parameter $m_{\psi}$ becomes small near the critical
point, the spin fractionalization cannot be ignored and thus,
should be taken into account appropriately. Neel fields and Dirac
spinons should be treated on equal footing. In order to solve the
Kondo-like spin coupling term $S_{M}$ in Eq. (1), we utilize the
$CP^{1}$ representation ${\vec n} =
\frac{1}{2}z^{\dagger}_{\sigma}{\vec
\tau}_{\sigma\sigma'}z_{\sigma'}$ or equally, ${\vec n}\cdot{\vec
\tau} = U\tau^{3}U^{\dagger}$, where $z_{\sigma}$ is a bosonic
spinon and $U$, its corresponding SU(2) matrix
$U= \left( \begin{array}{cc} z_{\uparrow} & -z_{\downarrow}^{\dagger} \\
z_{\downarrow} & z_{\uparrow}^{\dagger} \end{array} \right)$.
Inserting this $CP^{1}$ representation into Eq. (1), the Kondo
coupling term of $m_{\psi}\bar{\psi}_{\sigma}{\vec
\tau}_{\sigma\sigma'}\psi_{\sigma'}\cdot{\vec n}$ is written to be
${\cal L}_{K} = m_{\psi}\bar{\psi}_{\sigma}(U{\tau}^{3}
U^{\dagger})_{\sigma\sigma'}\psi_{\sigma'} $. Now this Kondo
coupling term can be easily solved by the following gauge
transformation \bqa \Psi_{\sigma} =
U^{\dagger}_{\sigma\sigma'}\psi_{\sigma'} , \eqa resulting in
${\cal L}_{K} = m_{\psi}{\bar
\Psi}_{\sigma}\tau^{3}_{\sigma\sigma'}\Psi_{\sigma'}$. Physical
meaning of this gauge transformation is clear. At the $QCP$
critical bosonic spinons $U^{\dagger}_{\sigma\sigma'}$ are
expected to hybridize with the fermionic spinons $\psi_{\sigma'}$
via the Kondo coupling, thus screening out their spin degrees of
freedom and resulting in the spinless fermions $\Psi_{\sigma}$.

Representing Eq. (1) in terms of the fractionalized fields
$U_{\sigma\sigma'}$ and $\Psi_{\sigma}$, we already mentioned that
the Kondo coupling term is completely solved to be $m_{\psi}{\bar
\Psi}_{\sigma}\tau^{3}_{\sigma\sigma'}\Psi_{\sigma'}$. In this
representation the scattering effect of fermionic spinons
$\psi_{\sigma}$ by spin fluctuations ${\vec n}$ appears in the
kinetic energy of spinons as
$\bar{\Psi}_{\sigma}\gamma_{\mu}(U^{\dagger}\partial_{\mu}U)_{\sigma\sigma'}\Psi_{\sigma'}$,
where currents of the spinless fermions couple to those of the
bosonic spinons, \bqa && Z = \int{DU}{D\Psi_{\sigma}}{Da_{\mu}}
e^{-S} , \nn && S = \int{d^3x} \Bigl[
\bar{\Psi}_{\sigma}\gamma_{\mu}([\partial_{\mu} -
ia_{\mu}]\delta_{\sigma\sigma'} +
[U^{\dagger}\partial_{\mu}U]_{\sigma\sigma'} )\Psi_{\sigma'} \nn
&& +
m_{\psi}\bar{\Psi}_{\sigma}{\tau}^{3}_{\sigma\sigma'}\Psi_{\sigma'}
+ \frac{1}{2e^2}|\partial\times{a}|^2 \nn && +
Tr\Bigl(\frac{Nm_{\psi}}{8\pi}|\partial_{\mu}(U\tau^3U^{\dagger})|^{2}\Bigr)
\Bigr] . \eqa This action has $U_{a}(1)\bigotimes{U}_{c}(1)$ local
gauge symmetry. The former $U_{a}(1)$ guarantees the invariance of
the action Eq. (4) under the gauge transformations of \bqa &&
\psi'_{\sigma} = e^{i\theta}\psi_{\sigma} \mbox{,   }\mbox{
}\mbox{   }\mbox{   }\mbox{   } \Psi'_{\sigma} =
e^{i\theta}\Psi_{\sigma} , \nn && U'_{\sigma\sigma'} =
U_{\sigma\sigma'} \mbox{,   }\mbox{   }\mbox{   }\mbox{   }\mbox{
} a'_{\mu} = a_{\mu} +
\partial_{\mu}\theta . \eqa This is nothing but the gauge symmetry
of the U(1) slave boson representation. On the other hand, the
latter $U_{c}(1)$ confirms the invariance of Eq. (4) under the
gauge transformations of \bqa && \psi'_{\sigma} = \psi_{\sigma'}
\mbox{, }\mbox{ }\mbox{ }\mbox{   }\mbox{   } \Psi'_{\sigma} =
[e^{i\vartheta\tau^{3}}]_{\sigma\sigma'}\Psi_{\sigma'} , \nn &&
U'_{\sigma\sigma'} =
U_{\sigma\alpha}[e^{-i\vartheta\tau^{3}}]_{\alpha\sigma'} \mbox{,
}\mbox{   }\mbox{   }\mbox{   }\mbox{   } a'_{\mu} = a_{\mu} .
\eqa {\it This new local gauge symmetry originates from the
$CP^{1}$ representation and implies that there should be a U(1)
gauge field corresponding to the gauge symmetry}. Indeed,
performing some standard algebra such as the Hubbard-Stratonovich
transformation, we can reach the following effective action \bqa
&& Z = \int{DU}{Dc_{\mu}}{D\Psi_{\sigma}}{Da_{\mu}} e^{-S} , \nn
&& S = \int{d^3x} \Bigl[
\bar{\Psi}_{\sigma}\gamma_{\mu}([\partial_{\mu} -
ia_{\mu}]\delta_{\sigma\sigma'} - ic_{\mu}\tau^{3}_{\sigma\sigma'}
)\Psi_{\sigma'} \nn && +
m_{\psi}\bar{\Psi}_{\sigma}{\tau}^{3}_{\sigma\sigma'}\Psi_{\sigma'}
-\frac{\pi}{Nm_{\psi}}|\bar{\Psi}_{\sigma}\gamma_{\mu}\tau^{3}_{\sigma\sigma'}\Psi_{\sigma'}|^{2}
+ \frac{1}{2e^2}|\partial\times{a}|^2 \nn && +
Tr\Bigl(\frac{Nm_{\psi}}{4\pi} |(\partial_{\mu} - ic_{\mu}\tau^{3}
)U^{\dagger}|^{2} \Bigr) + \frac{1}{2g^2}|\partial\times{c}|^2
\Bigr] . \eqa Here $c_{\mu}$ is a new compact U(1) gauge field
guaranteeing the $U_{c}(1)$ local gauge symmetry, where the gauge
field $c_{\mu}$ is transformed into $c'_{\mu} = c_{\mu} +
\partial_{\mu}\vartheta$ under the transformations in Eq. (6). $g$ is
an internal charge of the bosonic spinons
$U^{\dagger}_{\sigma\sigma'}$ and the spinless fermions
$\Psi_{\sigma}$. The spinless fermions have both internal charges,
$e$ and $g$. The kinetic energy of the SU(2) matrix $U^{\dagger}$
can be expressed into the familiar $CP^{1}$ representation,
$Tr\Bigl(\frac{Nm_{\psi}}{4\pi} |(\partial_{\mu} -
ic_{\mu}\tau^{3} )U^{\dagger}|^{2} \Bigr) = \frac{Nm_{\psi}}{2\pi}
|(\partial_{\mu} - ic_{\mu})z_{\sigma}|^{2}$. The Maxwell kinetic
energy of the gauge field $c_{\mu}$ can be safely introduced as
the contribution of high energy bosonic spinons because the
Maxwell gauge action is irrelevant in $(2+1)D$ in the
renormalization group ($RG$) sense and thus, does not affect
physics of the $QCP$.

We would like to emphasize that {\it Eq. (7) is just another
representation of Eq. (1)}. For this {\it exact} transformation to
be {\it physically} meaningful beyond a mathematical derivation,
there should exist the deconfined quantum criticality of the
$NL\sigma{M}$ allowing deconfined bosonic spinons at the $QCP$. In
other words, at least the $NL\sigma{M}$ in the effective field
theory Eq. (7) should be stable {\it in the $RG$ sense}. This is
the reason why we use the term, "based on the conjecture of the
deconfined quantum criticality," in the introduction. Fortunately,
this deconfined quantum criticality was shown to exist in Refs.
\cite{Senthil_deconfinement,Kim1} by an $RG$ analysis, as will be
discussed later. Based on the exact transformation and the
existence of the deconfined quantum criticality, we discuss the
$QCP$ between the $AF$ and the $ASL$. When the deconfinement of
bosonic spinons is not allowed owing to instanton effects of
compact U(1) gauge fields $c_{\mu}$, the effective field theory
Eq. (7) becomes unstable in the present decoupling scheme and
thus, another effective theory necessarily results. This indeed
happens away from the $QCP$, where the O(3) $NL\sigma{M}$ with the
Neel vectors and the $QED_3$ with the massless Dirac spinons and
slave boson U(1) gauge fields recover from Eq. (7) in the $AF$ and
in the $ASL$, respectively.

The gauge transformation Eq. (3) introduced for solving the Kondo
coupling term in Eq. (1) may be still suspected to be unnatural
although similar decoupling schemes have been utilized in the
context of the Kondo lattice
model\cite{Coleman_Kondo,Pepin_Kondo,Kim_Kondo2}. In this respect
it is necessary to understand the present methodology more deeply
by comparing this with other well studied ones. A good example is
a $d-wave$ BCS theory for superconductivity of high $T_c$
cuprates\cite{Fisher_Z2,Tesanovic}. In the context of $d-wave$
superconductivity the coupling term of
$|\Delta|e^{i\phi}c_{\uparrow}c_{\downarrow}$ between Cooper pairs
and electrons plays the same role as the Kondo coupling term of
$m_{\psi}{\vec n}\cdot{\bar \psi}_{\sigma}{\vec
\tau}_{\sigma\sigma'}\psi_{\sigma'}$ between spin fluctuations and
Dirac spinons in the context of antiferromagnetism. Here
$|\Delta|$ and $\phi$ are the amplitude and phase of Cooper pair
fields, analogous to $m_{\psi}$ and ${\vec n}$, respectively. In
order to solve this coupling term several kinds of gauge
transformations are introduced\cite{Fisher_Z2,Tesanovic}. In these
decoupling schemes critical phase fluctuations of Cooper pairs
screen out charge degrees of freedom of electrons, causing
electrically neutral but spinful electrons called "spinons". As a
result the phase factor disappears in the coupling term when it is
rewritten in terms of spinons. Instead, this coupling effect
appears as current-current interactions of neutral spinons and
phase fields of Cooper pairs in the kinetic energy of electrons.
Depending on the gauge transformations, $Z_2$\cite{Fisher_Z2} or
U(1)\cite{Tesanovic} gauge fields are obtained. {\it In this
respect the gauge transformation Eq. (3) naturally extends the
methodology of charge U(1) symmetry in the context of
superconductivity to that of spin SU(2) symmetry in the context of
antiferromagnetism}.

In order to convince critical physicists of the effective action
Eq. (7), in appendix D we show that Eq. (7) can be derived
directly from the antiferromagnetic Heisenberg Hamiltonian in a
standard fashion. We would like to suggest one good
reference\cite{Hubbard_Sigma} which derives a similar effective
action with Eq. (7) from the Hubbard Hamiltonian using the same
decoupling scheme.

\begin{table*}
\caption{Quantum phase transition from antiferromagnetism to
algebraic spin liquid}
\begin{tabular}{cccccccc}
\hline & Antiferromagnetism & Quantum Critical Point & Algebraic
Spin Liquid \nn & $m_{\psi} > m_{\psi}^{c}$ & $m_{\psi} =
m_{\psi}^{c}$ & $m_{\psi} < m_{\psi}^{c}$ \nn \hline
  Order Parameter & $<U^{\dagger}_{\sigma\sigma'}> \not= 0$
  ($<z_{\sigma}> \not= 0$) & & $<U^{\dagger}_{\sigma\sigma'}> = 0$ ($<z_{\sigma}> = 0$) \nn
  $g$ & $U_{\sigma\sigma'}\Psi_{\sigma'} \rightarrow \psi_{\sigma}$
  & $\psi_{\sigma} \rightarrow U_{\sigma\sigma'}\Psi_{\sigma'}$
  & $U_{\sigma\sigma'}\Psi_{\sigma'} \rightarrow \psi_{\sigma}$ \nn
  & Confinement of $g$ & Deconfinement of $g$ & Confinement of $g$
  \nn $e$ & $\psi_{\sigma}\tau^{\pm}_{\sigma\sigma'}\psi_{\sigma'} \rightarrow \pi^{\pm}$
  & ${\bar \Psi}_{\sigma}\Psi_{\sigma} \rightarrow {\bar \psi}_{\sigma}\psi_{\sigma}$
  & $\psi_{\sigma}$ \nn & Confinement of $e$ & Confinement of $e$
  & Deconfinement of $e$ \nn Elementary & $\pi^{\pm}$ & $U^{\dagger}_{\sigma\sigma'}$ ($z_{\sigma}$), $c_{\mu}$
  & $\psi_{\sigma}$, $a_{\mu}$ \nn Excitations & Antiferromagnons
  & Critical Bosonic Spinons & Critical Fermionic Spinons \nn
  \hline
\end{tabular}
\end{table*}

\subsection{Quantum Phase Transition from Antiferromagnetism to
Algebraic Spin Liquid}

Now we discuss the quantum phase transition between the $AF$ and
the $ASL$ based on Eq. (7). The critical value of the mass
parameter $m_{\psi}$ for the antiferro- to para- magnetic
transition is assumed to be $m_{\psi}^{c}$. In the case of
$m_{\psi} > m_{\psi}^{c}$ condensation of bosonic spinons occurs,
$<U^{\dagger}_{\sigma\sigma'}> \not= 0$ ($<z_{\sigma}> \not= 0$),
causing the $AF$. The spinon condensation leads the gauge field
$c_{\mu}$ to be massive (Anderson-Higgs mechanism). In the context
of gauge theories this phase corresponds to the Higgs-confinement
phase\cite{Fradkin,NaLee}, where the matter fields with internal
charge $g$ are confined to form gauge singlets. The condensed
bosonic spinons $U_{\sigma\sigma'}$ are confined with the spinless
fermions $\Psi_{\sigma'}$ and other bosonic spinons
$U^{\dagger}_{\sigma\sigma'}$ to make the fermionic spinons
$\psi_{\sigma}$ and the Neel order parameter fields ${\vec n}$,
respectively. Integrating over the massive U(1) gauge field
$c_{\mu}$ in Eq. (7), we obtain the following Lagrangian of ${\cal
L} = \bar{\psi}_{\sigma}\gamma_{\mu}(\partial_{\mu} -
ia_{\mu})\psi_{\sigma} + \frac{1}{2e^2}|\partial\times{a}|^2 +
m_{\psi}\bar{\psi}_{\sigma}{\vec
\tau}_{\sigma\sigma'}\psi_{\sigma'}\cdot{\vec n} +
\frac{Nm_{\psi}}{4\pi}|\partial_{\mu}{\vec n}|^{2}$, where the
local interactions between spinless fermions are cancelled. In the
$AF$ the Neel vectors align to the $z$ direction. Admitting
transverse fluctuations of the Neel vectors, we obtain an
effective field theory, ${\cal L}_{AF} =
\bar{\psi}_{n}\gamma_{\mu}(\partial_{\mu} - ia_{\mu})\psi_{n} +
m_{\psi}\bar{\psi}_{n}\tau^{3}_{nm}\psi_{m} +
\frac{1}{2e^2}|\partial\times{a}|^2 -
\frac{1}{2}m_{\psi}\bar{\psi}_{n}\tau^{3}_{nm}\psi_{m}|{\vec
\pi}|^{2} + m_{\psi}\bar{\psi}_{n}{\vec
\tau}_{nm}\psi_{m}\cdot{\vec \pi} + \frac{Nm_{\psi}}{4\pi}\Bigl(
|\partial_{\mu}{\vec \pi}|^{2} + ({\vec
\pi}\cdot\partial_{\mu}{\vec \pi})^{2} \Bigr)$, where we used
$n^{3} = \sqrt{1-|{\vec \pi}|^2} \approx 1 - \frac{1}{2}|{\vec
\pi}|^{2}$ as Eq. (2). The massive Dirac spinons should be
confined to form the antiferromagnons $\pi^{\pm}$. As a result the
conventional $AF$ described by Eq. (2) recovers from Eq. (7).

In our recent study\cite{Kim_zeromode} we proposed an interesting
possibility that massive Dirac spinons can appear to make broad
continuum spectrum at high energies in inelastic neutron
scattering\cite{Experiment}. The mechanism of spinon deconfinement
results from the existence of fermion zero modes in single
instanton potentials. Neel vectors can make a skyrmion
configuration around an instanton of the compact U(1) gauge field
$a_{\mu}$. Remarkably, in the instanton-skyrmion composite
potential the Dirac spinon is shown to have a zero mode. The
emergence of the fermion zero mode forbids the condensation of
instantons, resulting in the deconfinement of Dirac spinons in the
$AF$. Notice that the Dirac spinons are massive owing to
$S\chi{S}B$.

Approaching the $QCP$ ($m_{\psi} \rightarrow m_{\psi}^{c}$), the
critical bosonic spinons $z_{\sigma}$
($U^{\dagger}_{\sigma\sigma'}$) would be
deconfined\cite{Laughlin_deconfinement,Senthil_deconfinement,
Kim1,Ichinose_deconfinement} to exhibit broad continuum spectrum
in spin susceptibility. In Ref. \cite{Senthil_deconfinement} Berry
phase is shown to play a crucial role for the deconfinement of
bosonic spinons at the $QCP$. Berry phase gives rise to
destructive interference for instanton excitations of the
$c_{\mu}$ gauge fields. As a result only quadrupled instanton
excitations contribute to a partition function for the
$NL\sigma{M}$. These quadrupled instanton excitations are known to
be irrelevant at the $QCP$\cite{Senthil_deconfinement}. On the
other hand, in Refs.
\cite{Laughlin_deconfinement,Kim1,Ichinose_deconfinement} the
spinon deconfinement is claimed to occur even {\it in the absence
of Berry phase}. In these
studies\cite{Laughlin_deconfinement,Kim1,Ichinose_deconfinement}
the crucial point is the existence of quantum criticality itself.
Internal electric charges of spinons can be sufficiently screened
by critical fluctuations of spinons. The smaller internal electric
charges, the larger internal magnetic charges. As a result
instanton excitations can be irrelevant at the $QCP$. Especially,
in Ref. \cite{Kim1} the present author pointed out that the
$CP^{1}$ action of the $NL\sigma{M}$ in the easy plane limit has
two kinds of fixed points owing to an additional vortex gauge
field in the dual vortex representation. One is the charged fixed
point (inverted XY fixed point) and the other, the neutral fixed
point (XY fixed point). In Ref. \cite{Kim1} the inverted XY fixed
point is identified with the $QCP$ studied by Senthil et
al.\cite{Senthil_deconfinement}. The inverted XY fixed point is
shown to be unstable against instanton excitations and the
instanton excitations are proliferated. Since condensation of
instantons does not allow spinon unbinding, this is consistent
with the conclusion of Senthil and coworkers, that is, with the
absence of spinon deconfinement without a Berry phase term. At the
inverted XY critical point the spinon deconfinement can occur only
in the presence of Berry phase. On the other hand, the XY fixed
point is shown to remain stable against instanton excitations.
From the $RG$ analysis in Ref. \cite{Kim1} one sees that although
off criticality the instantons are relevant everywhere, they
become irrelevant at the XY $QCP$. This allows spinon
deconfinement, which is different from the conclusion by Senthil
et al.\cite{Senthil_deconfinement}, since this novel critical
point does not coincide with their one. The XY fixed point is,
instead, identified with the $QCP$ studied by Bernevig et
al.\cite{Laughlin_deconfinement}, thus showing that the
deconfinement of spinons takes place {\it even without the Berry
phase}. See Ref. \cite{Kim1} for details. We would like to mention
the result of Monte-Carlo simulation\cite{Ichinose_deconfinement}
supporting the existence of deconfined spinons at the $QCP$ of the
$O(3)$ $NL\sigma{M}$ without Berry phase. They claimed that
critical fluctuations of bosonic spinons result in a nonlocal
action of U(1) gauge fields and this contribution causes the
deconfinement of spinons\cite{Ichinose_deconfinement}.

Although the internal charge $g$ is deconfined owing to critical
fluctuations of the bosonic spinons, the spinless fermions
$\Psi_{\sigma}$ would not appear at the $QCP$. Remember that the
$\Psi_{\sigma}$ carries the internal charge $e$. The spinless
fermions would feel confining interactions mediated by U(1) gauge
fluctuations $a_{\mu}$ owing to their finite mass $m_{\psi}^{c}$
at the $QCP$. Ignoring antiferromagnetic spin fluctuations, i.e.,
${\vec n} = {\hat z}$ in the mean field level, one can find that
the antiferro- to para- magnetic transition is determined by the
self-consistent gap equation in Ref. \cite{CSB1} (appendix B). In
this case the spinon mass $m_{\psi}$ is zero at the critical
point. However, allowing spin fluctuations beyond the mean field
level, the critical value of the spinon mass for the magnetic
transition would not be zero. This can be justified by the $RG$
analysis of the O(3) $NL\sigma{M}$. Using the standard poor man's
scaling for the $S_{NL\sigma{M}}$ in Eq. (1), one obtains the well
known $RG$ equation for the spin stiffness $m_{\psi}$ in
$(2+1)D$\cite{Auerbach} \bqa && \frac{dm_{\psi}^{-1}}{dl} = -
m_{\psi}^{-1} + \lambda{{m}_{\psi}^{-1}}^{2} , \eqa where $l$ is a
scaling parameter and $\lambda$, a positive numerical constant.
This $RG$ equation yields three fixed points; $m_{\psi}
\rightarrow \infty$, $m_{\psi} \rightarrow 0$, and $m_{\psi}
\rightarrow m_{\psi}^{c} = \lambda$, corresponding to $AF$,
quantum disordered paramagnetism, and $QCP$, respectively. The
nonzero critical coupling in $m^{c}_{\psi}\bar{\psi}_{\sigma}{\vec
\tau}_{\sigma\sigma'}\psi_{\sigma'}\cdot{\vec n}$ at the $QCP$
seems to be consistent with that of the Kondo lattice
model\cite{Senthil_Kondo,Coleman_Kondo,Pepin_Kondo} which has a
similar structure with Eq. (7). Integration over the massive
fermions generates the Maxwell kinetic energy of the gauge field
$a_{\mu}$. In this case the well known classic result by
Polyakov\cite{Polyakov} can be applied, leading to confinement of
the charge $e$. The resulting critical field theory is obtained
from Eq. (7) to be ${\cal L}_{QCP} = \frac{Nm_{\psi}^{c}}{2\pi}
|(\partial_{\mu} - ic_{\mu})z_{\sigma}|^{2} +
\frac{1}{2g^2}|\partial\times{c}|^2$, where $c_{\mu}$ is a {\it
noncompact} U(1) gauge field. In appendix E, in order to confirm
this critical field theory we derive the similar critical theory
in a different way, where the Neel fields are not introduced.

In the case of $m_{\psi} < m_{\psi}^{c}$ the bosonic spinons are
gapped, $<U^{\dagger}_{\sigma\sigma'}> = 0$ ($<z_{\sigma}> = 0$).
In order to correctly describe the $ASL$, the massive bosonic
spinons should be confined with the spinless fermions to form the
fermionic spinons. In other words, the internal charge $g$ should
be confined via gauge fluctuations $c_{\mu}$. As the $RG$ Eq. (8)
shows $m_{\psi} \rightarrow 0$, the spinless fermions
$\Psi_{\sigma}$ would become massless in the quantum disordered
paramagnetism. One can think that the vanishing mass makes the
$\Psi_{\sigma}$ field critical, resulting in deconfinement of the
internal charge $g$ owing critical fluctuations of these fermions.
However, this guess is not correct. The vanishing fermion mass
gives rise to one problem that the strength of local interactions
goes to infinity in Eq. (7). In order to treat the infinitely
strong local interactions, we perform the Hubbard-Stratonovich
transformation in Eq. (7) to obtain $S = \int{d^3x} \Bigl[
\bar{\Psi}_{\sigma}\gamma_{\mu}([\partial_{\mu} -
ia_{\mu}]\delta_{\sigma\sigma'} - ic_{\mu}\tau^{3}_{\sigma\sigma'}
)\Psi_{\sigma'} + \frac{1}{2e^2}|\partial\times{a}|^2 +
\frac{1}{2{\bar g}^2}|\partial\times{c}|^2
-i\alpha_{\mu}\bar{\Psi}_{\sigma}\gamma_{\mu}\tau^{3}_{\sigma\sigma'}\Psi_{\sigma'}
\Bigr]$, where the gapped bosonic spinons are integrated out to
produce the Maxwell kinetic energy of the gauge field $c_{\mu}$
with a renormalized coupling strength ${\bar g}$, and
$\alpha_{\mu}$ is an auxiliary field to impose the local
interactions. In the above we utilized $m_{\psi} \rightarrow 0$.
Integration over the auxiliary field $\alpha_{\mu}$ results in the
local constraint,
$\bar{\Psi}_{\sigma}\gamma_{\mu}\tau^{3}_{\sigma\sigma'}\Psi_{\sigma'}
= 0$, indicating that the $\Psi_{\sigma}$ fields do not screen out
the internal charge $g$. Only the Maxwell kinetic energy is
available, causing confinement of the charge $g$\cite{Polyakov}.
The gapped bosonic spinons should be confined with the spinless
fermions to make the Dirac spinons. Since the effective field
theory Eq. (7) is unstable in the parameter range of $m_{\psi} <
m_{\psi}^{c}$, a new effective theory necessarily results.
Inserting $\Psi_{\sigma} = U_{\sigma\sigma'}\psi_{\sigma'}$ into
Eq. (7) under the constraint of
$\bar{\Psi}_{\sigma}\gamma_{\mu}\tau^{3}_{\sigma\sigma'}\Psi_{\sigma'}
= 0$ and integrating over the gapped bosonic spinons
$U^{\dagger}_{\sigma\sigma'}$, we can obtain the following
Lagrangian of ${\cal L}_{ASL} =
\bar{\psi}_{\sigma}\gamma_{\mu}(\partial_{\mu} -
ia_{\mu})\psi_{\sigma} + \frac{1}{2e^2}|\partial\times{a}|^2$.
Local interactions of spinon currents
$|\bar{\psi}_{\sigma}\gamma_{\mu}\psi_{\sigma}|^{2}$ would appear
in the above effective Lagrangian, but this term does not affect
low energy physics because it is irrelevant in $(2+1)D$ in the
$RG$ sense. As discussed earlier, the Dirac spinons with the
internal charge $e$ are deconfined to emerge at the charged fixed
point of the ${\cal L}_{ASL}$. We summarize the quantum phase
transition from the $AF$ to the $ASL$ in Table I.

\section{Summary}

In summary, we proposed the fractionalization of the Dirac spinons
into the bosonic spinons and spinless fermions near the quantum
critical point between the antiferromagnetism and the algebraic
spin liquid (Table I). Based on this conjecture we constructed the
"mother" critical field theory Eq. (7) and showed that the
antiferromagnetism and the algebraic spin liquid can successfully
recover from Eq. (7).

\section{Acknowledgement}

K.-S. Kim would like to thank Dr. A. Tanaka for pointing out his
misunderstanding in the early stage of the present work. K.-S. Kim
also thanks Dr. Sung-Sik Lee, Dr. Park, Tae-Sun for helpful
discussions and Dr. Sreedhar B. Dutta for his reading of this
manuscript.

\appendix
\section{}

In appendix A we briefly sketch how to obtain the $S_{ASL}$ in Eq.
(1) from the antiferromagnetic Heisenberg model on two dimensional
square lattices, $H_{J} =
J\sum_{<i,j>}\vec{S}_{i}\cdot\vec{S}_{j}$ with $J>0$. Inserting
the following spinon representation for spin, $\vec{S}_{i} =
\frac{1}{2}f^{\dagger}_{i\sigma}\vec{\tau}_{\sigma\sigma'}f_{i\sigma'}$
into the above Heisenberg model, and performing the standard
Hubbard-Stratonovich transformation for an exchange interaction
channel, we obtain an effective one body Hamiltonian for fermions
coupled to an order parameter, $H_{eff} =
-J\sum_{<i,j>}f_{i\sigma}^{\dagger}\chi_{ij}f_{j\sigma} - h.c.$.
Here $f_{i\sigma}$ is a fermionic spinon with spin $\sigma =
\uparrow, \downarrow$, and $\chi_{ij}$ is an auxiliary field
called a hopping order parameter. Notice that the hopping order
parameter $\chi_{ij}$ is a complex number defined on links $ij$.
Thus, it can be decomposed into $\chi_{ij} =
|\chi_{ij}|e^{i\theta_{ij}}$, where $|\chi_{ij}|$ and
$\theta_{ij}$ are the amplitude and phase of the hopping order
parameter, respectively. Inserting this representation for the
$\chi_{ij}$ into the effective Hamiltonian, we obtain $H_{eff} =
-J\sum_{<i,j>}|\chi_{ij}|f_{i\sigma}^{\dagger}e^{i\theta_{ij}}f_{j\sigma}
- h.c.$ We can easily see that this effective Hamiltonian has an
internal U(1) gauge symmetry, $H'_{eff}[f'_{i\sigma},\theta'_{ij}]
= H_{eff}[f_{i\sigma},\theta_{ij}]$ under the following U(1) phase
transformations, $f'_{i\sigma} = e^{i\phi_{i}}f_{i\sigma}$ and
$\theta'_{ij} = \theta_{ij} + \phi_{i} - \phi_{j}$. This implies
that the phase field $\theta_{ij}$ of the hopping order parameter
plays the same role as a U(1) gauge field $a_{ij}$. When a spinon
hops on lattices, it obtains an Aharnov-Bohm phase owing to the
U(1) gauge field $a_{ij}$. It is known that a stable mean field
phase is a $\pi$ flux state if an antiferromagnetic order is not
taken into account\cite{Review,DonKim_QED3}. This means that a
spinon gains the phase of $\pi$ when it turns around one
plaquette. In the $\pi$ flux phase low energy elementary
excitations are massless Dirac spinons near nodal points showing
gapless Dirac spectrum and U(1) gauge
fluctuations\cite{Review,DonKim_QED3}. In the low energy limit the
amplitude $|\chi_{ij}|$ is frozen to $|\chi_{ij}| =
J|<f_{j\sigma}^{\dagger}f_{i\sigma}>|$. A resulting effective
field theory for this possible quantum paramagnetism of the
antiferromagnetic Heisenberg model is $QED_3$ in terms of massless
Dirac spinons interacting via compact U(1) gauge fields, $S_{ASL}$
in Eq. (1).

In the $S_{ASL}$ $\psi_{\sigma} = \left(
\begin{array}{c} \chi^{+}_{\sigma} \\  \chi^{-}_{\sigma} \end{array} \right)$
is a four component massless Dirac fermion, where $\sigma = 1, 2$
represents its SU(2) spin ($\uparrow, \downarrow$) and $\pm$
denote the nodal points of $(\pi/2,\pm\pi/2)$ in momentum space.
Usually, SU(N) quantum antiferromagnets are considered by
generalizing the spin components $\sigma = 1, 2$ into $\sigma = 1,
2, ..., N$. The two component spinors $\chi^{\pm}_{\sigma}$ are
given by $\chi^{+}_{1} = \left(
\begin{array}{c} f_{\uparrow{1e}} \\ f_{\uparrow{1o}} \end{array}
\right)$, $\chi^{-}_{1} = \left( \begin{array}{c} f_{\uparrow{2o}}
\\ f_{\uparrow{2e}} \end{array} \right)$,
$\chi^{+}_{2} = \left( \begin{array}{c} f_{\downarrow{1e}}
\\ f_{\downarrow{1o}} \end{array} \right)$, and
$\chi^{-}_{2} = \left( \begin{array}{c} f_{\downarrow{2o}} \\
f_{\downarrow{2e}} \end{array} \right)$, respectively. In the
spinon field $f_{abc}$ $a = \uparrow, \downarrow$ represents its
SU(2) spin, $b = 1, 2$, the nodal points ($+, -$), and $c = e, o$,
even and odd sites, respectively\cite{DonKim_QED3}. Dirac matrices
$\gamma_{\mu}$ are given by $\gamma_{0} = \left( \begin{array}{cc}
\sigma_{3} & 0 \\ 0 & -\sigma_{3} \end{array} \right)$,
$\gamma_{1} = \left( \begin{array}{cc} \sigma_{2} & 0 \\ 0 &
-\sigma_{2} \end{array} \right)$, and $\gamma_{2} = \left(
\begin{array}{cc} \sigma_{1} & 0 \\ 0 & -\sigma_{1} \end{array}
\right)$, respectively, where they satisfy the Clifford algebra
$[\gamma_{\mu},\gamma_{\nu}]_{+} =
2\delta_{\mu\nu}$\cite{DonKim_QED3}.

\section{}

In appendix B, for completeness of this paper we briefly sketch
how we obtain the dynamically generated spinon mass $m_{\psi}$ in
$S_{M}$ in Eq. (1). A single spinon propagator is given by
$G^{-1}(k) = G_{0}^{-1}(k) - \Sigma(k)$, where $G_{0}^{-1}(k) =
i\gamma_{\mu}k_{\mu}$ is the inverse of a bare spinon propagator,
and $\Sigma(k)$, a spinon self-energy resulting from long range
gauge interactions. The spinon self-energy is determined by the
self-consistent gap equation, $\Sigma(k) =
\int\frac{d^3q}{(2\pi)^{3}}Tr[\gamma_{\mu}G(k-q)\gamma_{\nu}D_{\mu\nu}(q)]$,
where $D_{\mu\nu}(q)$ is a renormalized propagator of the U(1)
gauge field $a_{\mu}$ due to particle-hole excitations of massless
Dirac fermions. The self-energy can be written as $\Sigma(k) = -
m_{\psi}(k)\tau^{3}$ for staggered
magnetization\cite{DonKim_QED3}. Inserting this representation
into the above self-consistent gap equation, we obtain the
following expression for the spinon mass, $m_{\psi}(p) =
\int\frac{d^3k}{(2\pi)^{3}}\frac{\gamma_{\mu}m_{\psi}(k)\gamma_{\nu}}{k^{2}
+ m_{\psi}^{2}(k)}D_{\mu\nu}(p-k)$. The renormalized gauge
propagator $D_{\mu\nu}(q)$ is obtained to be $D_{\mu\nu}(q)
\approx \Pi_{\mu\nu}^{-1}(q) = \frac{8}{Nq}\Bigl(\delta_{\mu\nu} -
\frac{q_{\mu}q_{\nu}}{q^2}\Bigr)$ in the Lorentz
gauge\cite{DonKim_QED3}, where $\Pi_{\mu\nu}(q) = -
N\int\frac{d^3k}{(2\pi)^{3}}Tr[\gamma_{\mu}G_{0}(k)\gamma_{\nu}G_{0}(k-q)]$
is the polarization function of massless Dirac fermions in the
$1/N$ approximation. Inserting this gauge propagator into the gap
equation and performing an angular integration, one can find
$m_{\psi}(p) =
\frac{4}{N\pi^2p}\int_{0}^{\Lambda}dk\frac{km_{\psi}(k)}{k^{2} +
m_{\psi}^{2}(k)}(k+p-|k-p|)$, where $\Lambda$ is a momentum
cutoff\cite{DonKim_QED3}. This integral expression is equivalent
to the differential equation,
$\frac{d}{dp}\Bigl(p^{2}\frac{dm_{\psi}(p)}{dp}\Bigr) = -
\frac{8}{\pi^2N}\frac{p^2m_{\psi}(p)}{p^2+m_{\psi}^2(p)}$ with
boundary conditions, $\Lambda\frac{dm_{\psi}(p)}{dp}_{p = \Lambda}
+ m_{\psi}(\Lambda) = 0$ and $0 \leq m_{\psi}(0) <
\infty$\cite{CSB1,DonKim_QED3}. In Ref. \cite{CSB1} this equation
is well analyzed in detail. Its solution is given by $m_{\psi}
\approx e^{2}exp[-2\pi/\sqrt{N_c/N - 1}]$ in the case of $N <
N_c$\cite{CSB1,Herbut}.

\section{}

In appendix C we discuss how the O(3) $NL\sigma{M}$ can be derived
from the effective action $S_{ASL} + S_{M}$ in Eq. (1).
Integration over the Dirac fermions in $S_{ASL} + S_{M}$ results
in the following expression \bqa && S_{NL\sigma{M}} = -
ln\Bigl[\int{D\psi_\sigma}
exp\{-\int{d^3x}\Bigl(\bar{\psi}_{\sigma}\gamma_{\mu}(\partial_{\mu}
- ia_{\mu})\psi_{\sigma} \nn && +
m_{\psi}\bar{\psi}_{\sigma}({\vec n}\cdot{\vec
\tau}_{\sigma\sigma'})\psi_{\sigma'} \Bigr)\}\Bigr] \nn && = -
Nln\det\Bigl[\gamma_{\mu}(\partial_{\mu} - ia_{\mu}) +
m_{\psi}{\vec n}\cdot{\vec \tau} \Bigr] \nn && \approx -
\frac{N}{2}ln\det(-\partial^{2} + m_{\psi}^{2}) \nn && +
\int{d^3x}\Bigl( \frac{Nm_{\psi}}{4\pi}|\partial_{\mu}{\vec
n}|^{2} + \frac{N}{12\pi{m}_{\psi}}|\partial\times{a}|^{2} \Bigr)
. \eqa The second term in the last line is well derived in Ref.
\cite{Dmitri}, \bqa && - \frac{N}{2}ln\det\Bigl[1 -
\frac{m_{\psi}\gamma_{\mu}\partial_{\mu}({\vec n}\cdot{\vec
\tau})}{-\partial^{2} + m_{\psi}^{2}}\Bigr] \nn && = -
\frac{N}{2}Tr\int{d^3x}\langle{x}|ln\Bigl[1 -
\frac{m_{\psi}\gamma_{\mu}\partial_{\mu}({\vec n}\cdot{\vec
\tau})}{-\partial^{2} + m_{\psi}^{2}}\Bigr]|x\rangle \nn && = -
\frac{N}{2} \int{d^3x}\int\frac{d^3k}{(2\pi)^{3}} \nn &&
e^{-ikx}Trln\Bigl[1 -
\frac{m_{\psi}\gamma_{\mu}\partial_{\mu}({\vec n}\cdot{\vec
\tau})}{-\partial^{2} + m_{\psi}^{2}}\Bigr]e^{ikx} \nn && = -
\frac{N}{2} \int{d^3x}\int\frac{d^3k}{(2\pi)^{3}} \nn &&
Trln\Bigl[1 - \frac{m_{\psi}\gamma_{\mu}\partial_{\mu}({\vec
n}\cdot{\vec \tau})}{k^{2} + m_{\psi}^{2} -
2ik_{\mu}\partial_{\mu} -
\partial^{2}}\Bigr] \nn && \approx \int{d^3x}
\frac{N}{4}\int\frac{d^3k}{(2\pi)^3}
Tr\Bigl(\frac{m_{\psi}\gamma_{\mu}\partial_{\mu}({\vec
n}\cdot{\vec \tau})}{k^{2} + m_{\psi}^{2}}\Bigr)^{2} \nn && =
\int{d^3x} \frac{Nm_{\psi}}{4\pi}|\partial_{\mu}{\vec n}|^{2} .
\eqa In the above $Tr$ stands for not a functional but a usual
matrix trace for both flavor (spin) and spinor indices. In going
from the third to the fourth line we have dragged the factor
$e^{ikx}$ through the operator, thus shifting all differential
operators $\partial_{\mu} \rightarrow
\partial_{\mu} + ik_{\mu}$\cite{Dmitri}. Expanding the argument of the
logarithmic term in powers of $\partial_{\mu}{\vec n}$ and of
$2ik_{\mu}\partial_{\mu} + \partial^{2}$, one can easily obtain
the expression in the fifth line. Performing the momentum
integration, we obtain an effective spin stiffness proportional to
the mass parameter $m_{\psi}$. This implies that the rigidity of
fluctuations in the Neel field is controlled by the mass parameter
$m_{\psi}$ of Dirac spinons. Eq. (C2) is nothing but the O(3)
$NL\sigma{M}$ describing a collinear spin order. Note that higher
order derivative terms in the gradient expansion are irrelevant in
$(2+1)D$ in the $RG$ sense.

Next, we sketch the derivation of the Maxwell gauge action.
Expanding the argument of the logarithmic term in Eq. (C1) to the
second order of the gauge field $a_{\mu}$, we obtain $S_{gauge} =
\int\frac{d^3q}{(2\pi)^3}\frac{1}{2}a_{\mu}(q)\Pi_{\mu\nu}(q)a_{\nu}(-q)$,
where the fermion polarization function $\Pi_{\mu\nu}(q)$ is given
by $\Pi_{\mu\nu}(q) = - N
\int\frac{d^3k}{(2\pi)^3}Tr[\gamma_{\mu}G(k+q)\gamma_{\nu}G(k)]$
with the single spinon propagator $G(k) = [i\gamma_{\mu}k_{\mu} +
m_{\psi}{\vec n}\cdot{\vec \tau}]^{-1}$. Utilizing the Feynman
identity and trace identity of Dirac gamma matrices, one can
obtain the following expression for the polarization function,
$\Pi_{\mu\nu}(q) =
2N(TrI)\frac{\Gamma(2-D/2)}{(4\pi)^{D/2}}(q^{2}\delta_{\mu\nu} -
q_{\mu}q_{\nu})\int_{0}^{1}dx(1-x)x(m_{\psi}^2+q^2x(1-x))^{D/2-2}
= \frac{(TrI)N}{4\pi}(q^{2}\delta_{\mu\nu} -
q_{\mu}q_{\nu})\Bigl(\frac{m_{\psi}}{2q^2} +
\frac{q^2-4m_{\psi}^2}{4q^3}\sin^{-1}\Bigl(\frac{q}{\sqrt{4m_{\psi}^2+q^2}}\Bigr)\Bigr)$\cite{DonKim_QED3}.
This leads to the Maxwell gauge action in Eq. (C1).

We should comment the reason why imaginary terms do not arise in
the present $NL\sigma{M}$ because some previous studies have shown
the emergence of imaginary
terms\cite{Abanov_sigma,Chiral_anomaly}. Following the evaluations
in Ref. \cite{Abanov_sigma}, one can obtain two imaginary terms
{\it in the irreducible representation of gamma matrices}; one is
a coupling term $ia_{\mu}J_{\mu}$ between topologically nontrivial
fermionic currents $J_{\mu} =
\frac{1}{8\pi}\epsilon_{\mu\nu\lambda}
\epsilon_{\alpha\beta\gamma} n^{\alpha}\partial_{\nu}n^{\beta}
\partial_{\lambda}n^{\gamma}$ and U(1) gauge fields $a_{\mu}$, and
the other, a geometrical phase $iN\pi\Gamma[{\vec
n}]$\cite{Abanov_sigma,Chiral_anomaly}. {\it The key point is the
representation of Dirac gamma matrices}\cite{A_Tanaka}. Here, we
utilized four-by-four Dirac matrices by combining the two nodal
points. Note that the signs of Pauli matrices in the Dirac gamma
matrices are opposite for the nodal points $\pm$. This fact
results in cancellation of the imaginary terms. The first
imaginary term can be considered from a variation of the
logarithmic term in Eq. (C1) with respect to the gauge field
$a_{\mu}$ \bqa &&
iNTr\Bigl[\gamma_{\mu}\delta{a}_{\mu}\frac{1}{\gamma_{\mu}(\partial_{\mu}
- ia_{\mu}) + m_{\psi}({\vec n}\cdot{\vec \tau})} \Bigr] \nn && =
iNTr\Bigl[\gamma_{\mu}\delta{a}_{\mu}\frac{-\gamma_{\mu}(\partial_{\mu}
- ia_{\mu}) + m_{\psi}({\vec n}\cdot{\vec
\tau})}{-[\gamma_{\mu}(\partial_{\mu} - ia_{\mu})]^{2} +
m_{\psi}^{2} - m_{\psi}\gamma_{\mu}\partial_{\mu}({\vec
n}\cdot{\vec \tau})} \Bigr] \nn && \approx
iNTr\Bigl[\gamma_{\mu}\delta{a}_{\mu}\frac{m_{\psi}({\vec
n}\cdot{\vec \tau})}{-\partial^{2} +
m_{\psi}^{2}}\Bigl(\frac{m_{\psi}\gamma_{\mu}\partial_{\mu}({\vec
n}\cdot{\vec \tau})}{-\partial^{2} + m_{\psi}^{2}}\Bigr)^{2}\Bigr]
. \eqa In this expression the key point is the triple product of
Dirac matrices, $\gamma_{\mu}\gamma_{\nu}\gamma_{\lambda}$ in the
last line. In the irreducible representation of gamma matrices,
i.e., Pauli matrices, this contribution is nonzero, leading to
$\epsilon_{\mu\nu\lambda}$. As a result the imaginary term of
$i{a}_{\mu}\epsilon_{\mu\nu\lambda} \epsilon_{\alpha\beta\gamma}
n^{\alpha}\partial_{\nu}n^{\beta}
\partial_{\lambda}n^{\gamma}$ can
be obtained\cite{Abanov_sigma,Chiral_anomaly}. Another
$\epsilon_{\alpha\beta\gamma}$ associated with the Neel vectors
appears from the triple product of Pauli matrices,
$\tau_{\alpha}\tau_{\beta}\tau_{\gamma}$. On the other hand, in
the present representation of Dirac matrices the contribution of
the $+$ nodal point leads to $+\epsilon_{\mu\nu\lambda}$ while
that of the $-$ nodal point, $-\epsilon_{\mu\nu\lambda}$. Thus,
{\it these two contributions are exactly cancelled}. The
geometrical phase term, considered from a variation of the
logarithmic term in Eq. (C1) with respect to the Neel field ${\vec
n}$\cite{Abanov_sigma,Chiral_anomaly}, \bqa && -
NTr\Bigl[m_{\psi}\delta({\vec n}\cdot{\vec
\tau})\frac{1}{\gamma_{\mu}\partial_{\mu} + m_{\psi}({\vec
n}\cdot{\vec \tau})} \Bigr] \nn && = -
NImTr\Bigl[m_{\psi}\delta({\vec n}\cdot{\vec
\tau})\frac{-\gamma_{\mu}\partial_{\mu} + m_{\psi}({\vec
n}\cdot{\vec \tau})}{-\partial^{2} + m_{\psi}^{2} -
m_{\psi}\gamma_{\mu}\partial_{\mu}({\vec n}\cdot{\vec \tau})}
\Bigr] \nn && \approx NTr\Bigl[m_{\psi}\delta({\vec n}\cdot{\vec
\tau})\frac{m_{\psi}({\vec n}\cdot{\vec \tau})}{-\partial^{2} +
m_{\psi}^{2}}\Bigl(\frac{m_{\psi}\gamma_{\mu}\partial_{\mu}({\vec
n}\cdot{\vec \tau})}{-\partial^{2} + m_{\psi}^{2}}\Bigr)^{3}\Bigr]
, \eqa is also exactly zero owing to the same reason. {\it Another
way to say this is that the signs of mass terms for the Dirac
fermions ($\chi_{n}^{+}$ and $\chi_{n}^{-}$) at the two Dirac
nodes ($+$ and $-$) are opposite, resulting in cancellation of the
parity anomaly}\cite{A_Tanaka}. If we fix the Neel vector in the
$z$ direction (${\vec n} = {\hat z}$), we can see {\it the
opposite signs} explicitly from $m_{\psi}
\bar{\psi}_{n}\tau^{z}_{nm}\psi_{m} = m_{\psi}
{\psi}^{\dagger}_{n}\gamma_{0}\tau^{z}_{nm}\psi_{m} =
m_{\psi}\chi^{+}_{n}\sigma_{z}\tau^{z}_{nm}\chi^{+}_{m} -
m_{\psi}\chi^{-}_{n}\sigma_{z}\tau^{z}_{nm}\chi^{-}_{m} =
m_{\psi}{\bar\chi}^{+}_{n}\tau^{z}_{nm}\chi^{+}_{m} -
m_{\psi}{\bar\chi}^{-}_{n}\tau^{z}_{nm}\chi^{-}_{m}$. Both massive
Dirac fermions ($\chi_{n}^{\pm}$) contribute to the imaginary
terms, respectively. However, the signs of the imaginary terms are
opposite and thus, the cancellation occurs. As a result the
imaginary terms do not appear in the present $NL\sigma{M}$. This
was already discussed in Refs. \cite{Zou,Wen}.

It is possible that the mass terms have the {\it same} signs.
Considering the two gamma matrices of $\gamma_{4} = \left(
\begin{array}{cc} 0 & I \\ I & 0 \end{array} \right)$ and $\gamma_{5} = \left(
\begin{array}{cc} 0 & I \\ - I & 0 \end{array} \right)$\cite{DonKim_QED3}, we can obtain
the following mass terms with the same signs, ${\cal
\tilde{L}}_{M} =
m_{\psi}\bar{\psi}_{n}\gamma_{4}\gamma_{5}\tau^{z}\psi_{m} = -
m_{\psi}{\bar\chi}^{+}_{n}\tau^{z}_{nm}\chi^{+}_{m} -
m_{\psi}{\bar\chi}^{-}_{n}\tau^{z}_{nm}\chi^{-}_{m}$. These mass
terms can arise from the $S_{ASL}$ in Eq. (1) via $S\chi{S}B$
because the $ASL$ has the enlarged
symmetry\cite{Tanaka_sigma,Hermele_sigma}, as discussed earlier.
In this case the cancellation dose not occur and thus, the
imaginary terms necessarily arise. This $AF$ would not be
conventional since it breaks not only time reversal symmetry but
also parity symmetry. When this $AF$ disappears via strong quantum
fluctuations, its corresponding quantum disordered paramagnet is
expected to be the {\it chiral spin
liquid}\cite{Tanaka_sigma,Zou,Wen}. In the present paper we did
not discuss the quantum phase transition from this anomalous
antiferromagnetism to the chiral spin liquid.

\section{}

In appendix D we briefly sketch a direct derivation of Eq. (7)
from the antiferromagnetic Heisenberg Hamiltonian $H_{J} =
J\sum_{<i,j>}{\vec S}_{i}\cdot{\vec S}_{j}$ at half filling.
Inserting the U(1) slave boson representation of the spin operator
${\vec S}_{i} = \frac{1}{2}f_{i\sigma}^{\dagger}{\vec
\tau}_{\sigma\sigma'}f_{i\sigma'}$ into the Heisenberg
Hamiltonian, we obtain $H_{J} =
\frac{J}{4}\sum_{<i,j>}(f_{i\sigma}^{\dagger}{\vec
\tau}_{\sigma\sigma'}f_{i\sigma'})\cdot(f_{j\alpha}^{\dagger}{\vec
\tau}_{\alpha\alpha'}f_{j\alpha'})$. These two body interactions
of spinons can be decoupled into the Hartree (direct), Fock
(exchange), Bogoliubov (pairing) channels respectively,
$\frac{J}{4}(f_{i\sigma}^{\dagger}{\vec
\tau}_{\sigma\sigma'}f_{i\sigma'})\cdot(f_{j\alpha}^{\dagger}{\vec
\tau}_{\alpha\alpha'}f_{j\alpha'}) = -
\frac{J}{8}[(f_{i\sigma}^{\dagger}{\vec
\tau}_{\sigma\sigma'}f_{i\sigma'})\cdot(f_{j\alpha}^{\dagger}{\vec
\tau}_{\alpha\alpha'}f_{j\alpha'}) +
(f_{i\sigma}^{\dagger}f_{i\sigma})(f_{j\sigma}^{\dagger}f_{j\sigma})]
-
\frac{J}{4}[(f_{i\sigma}^{\dagger}f_{j\sigma})(f_{j\sigma}^{\dagger}f_{i\sigma})
- f_{i\sigma}^{\dagger}f_{i\sigma}] -
\frac{J}{2}(f_{i\downarrow}^{\dagger}f_{j\uparrow}^{\dagger} -
f_{i\uparrow}^{\dagger}f_{j\downarrow}^{\dagger})(f_{j\uparrow}f_{i\downarrow}
- f_{j\downarrow}f_{i\uparrow}) + \frac{J}{4}$\cite{Sung_Sik}.
Performing the Hubbard-Stratonovich transformation for each
interaction channel, we obtain the following effective Lagrangian
\bqa && L_{eff} = \sum_{i}f_{i\sigma}^{\dagger}(\partial_{\tau} -
ia_{i\tau} - \mu)f_{i\sigma} -
\sum_{<i,j>}(f_{i\sigma}^{\dagger}\chi_{ij}f_{j\sigma} + h.c.) \nn
&& - \sum_{<i,j>}f_{i\sigma}^{\dagger}{\vec m}_{i}\cdot{\vec
\tau}_{\sigma\sigma'}f_{i\sigma'} -
\sum_{<i,j>}\rho_{i}f_{i\sigma}^{\dagger}f_{i\sigma} \nn && -
\sum_{<i,j>}[\Delta_{ij}(f_{i\downarrow}^{\dagger}f_{j\uparrow}^{\dagger}
- f_{i\uparrow}^{\dagger}f_{j\downarrow}^{\dagger}) + h.c.] \nn &&
+ \frac{1}{J}\sum_{ij}(|\chi_{ij}|^{2} +
\frac{1}{2}|\Delta_{ij}|^{2}) + \frac{2}{J}\sum_{i}(|{\vec
m}_{i}|^{2} + \rho_{i}^{2}) . \eqa Here $\chi_{ij}$ is an exchange
(hopping) order parameter, ${\vec m}_{i}$, a spin density wave
order parameter, $\rho_{i}$, a charge density wave order
parameter, and $\Delta_{ij}$, a pairing order parameter.
$a_{i\tau}$ is a Lagrange multiplier field to impose the single
occupancy constraint of $f_{i\sigma}^{\dagger}f_{i\sigma} = 1$,
and $\mu$, a spinon chemical potential formally introduced. In the
following we do not consider the pairing channel. Instead, we
focus our attention on the spinon exchange and magnetization
associated with the $ASL$ and $AF$, respectively. The ansatz of an
antiferromagnetic order leads us to the saddle point equations of
the three order parameters, $\chi_{ij} =
J\langle{f}_{j\sigma}^{\dagger}f_{i\sigma}\rangle$, ${\vec m}_{i}
=
\frac{J}{4}(-1)^{i}\langle{f}_{i\sigma}^{\dagger}\tau^{3}_{\sigma\sigma'}f_{i\sigma'}\rangle{\hat
z}$, and $\rho_{i} =
\frac{J}{4}\langle{f}_{i\sigma}^{\dagger}f_{i\sigma}\rangle$. At
half filling the saddle point value of the charge density wave
order parameter is given by $\rho_{i} = J/4$. Shifting the
$a_{i\tau}$ field into $ia_{i\tau} \rightarrow ia_{i\tau} - \mu -
J/4$ in Eq. (D1) leads to the following effective Lagrangian \bqa
&& L_{eff} = \sum_{i}f_{i\sigma}^{\dagger}(\partial_{\tau} -
ia_{i\tau})f_{i\sigma} -
\chi_{0}\sum_{<i,j>}(f_{i\sigma}^{\dagger}e^{ia_{ij}}f_{j\sigma} +
h.c.) \nn && - m\sum_{<i,j>}f_{i\sigma}^{\dagger}{\vec
\Omega}_{i}\cdot{\vec \tau}_{\sigma\sigma'}f_{i\sigma'} . \eqa
Here $\chi_0 = |\chi_{ij}|$ and $m = {\vec m}_{i}\cdot{\hat z}$
are amplitudes of the order parameters, assumed to be uniform in
space and time. $a_{ij}$ is the phase of $\chi_{ij}$ and ${\vec
\Omega}_{i}$, that of ${\vec m}_{i}$.

In order to derive Eq. (7) from the above effective Lagrangian Eq.
(D2), one can use the Haldane mapping of ${\vec \Omega}_{i} =
(-1)^{i}{\vec n}_{i}\sqrt{1-{\vec L}_{i}^{2}} + {\vec L}_{i}$,
where ${\vec n}_{i}$ is a unitary vector representing the Neel
field, and ${\vec L}_{i}$, the canting vector orthogonal to ${\vec
n}_{i}$\cite{Nagaosa,Auerbach}. Using the $CP^{1}$ representation
of ${\vec n}\cdot{\vec \tau} = U\tau^{3}U^{\dagger}$ and
introducing the gauge transformation of $d_{\sigma} =
U^{\dagger}_{\sigma\sigma'}f_{\sigma'}$ in Eq. (D2), one reaches
the following Lagrangian \bqa && L_{eff} =
\sum_{i}d_{i\sigma}^{\dagger}([\partial_{\tau} -
ia_{i\tau}]\delta_{\sigma\sigma'} -
[U^{\dagger}\partial_{\tau}U]_{\sigma\sigma'})d_{i\sigma'} \nn &&
-
\chi_{0}\sum_{<i,j>}(d_{i\sigma}^{\dagger}U^{\dagger}_{i\sigma\alpha}e^{ia_{ij}}U_{j\alpha\sigma'}d_{j\sigma'}
+ h.c.) \nn && -
m\sum_{i}d_{i\sigma}^{\dagger}[(-1)^{i}\tau^{3}_{\sigma\sigma'}\sqrt{1-{\vec
L}_{i}^{2}} + {\vec \tau}_{\sigma\sigma'}\cdot{\vec
L}_{i}]d_{i\sigma'} . \eqa Integrating over the spinless fermions
$d_{\sigma}$, expanding the resulting fermion determinant to the
second order of $U^{\dagger}\partial_{\mu}U$ and $L_{\mu}$, and
integrating out the canting fields $L_{\mu}$, one can finally
obtain the O(3) $NL\sigma{M}$. This procedure is well derived in
Ref. \cite{Hubbard_Sigma}. Combining the spinless fermions with
this $NL\sigma{M}$ in order to examine the magnetic $QCP$, and
performing the Hubbard-Stratonovich transformation used in section
III-A, one can find an effective action with an emergent U(1)
gauge field $c_{\mu}$ similar to Eq. (7) in the continuum limit.

We would like to comment how the present study can be expanded to
the case of geometrically frustrated lattices. Remember that in
the present paper a collinear spin order is considered to result
in the O(3) $NL\sigma{M}$ in Eq. (1). In the frustrated
antiferromagnets non-collinear spin orders are expected and thus,
the present $NL\sigma{M}$ cannot be applied. {\it In this case the
unitary vector ${\vec \Omega}$ in Eq. (D2) should be generalized
to impose non-collinear spin orders}. This requires a different
kind of Haldane mapping from the above. In this case a
paramagnetic phase resulting from each non-collinear spin order is
an interesting problem. Remember that in appendix C we mentioned
an exciting quantum phase transition from an anomalous
antiferromagnetism to the chiral spin liquid. These two
interesting studies remain as future works.

\section{}

In order to confirm the present description of the $QCP$, we
derive the similar critical field theory, but in a totally
different way where the Neel order parameter fields are not
introduced. We consider the following effective Hamiltonian
(appendix A and D) \bqa && H_{eff} = -
\chi_{0}\sum_{<i,j>}(f_{i\sigma}^{\dagger}e^{ia_{ij}}f_{j\sigma} +
h.c.) \nn && - m\sum_{<i,j>}(-1)^{i}f_{i\sigma}^{\dagger}{\vec
\tau}^{3}_{\sigma\sigma'}f_{i\sigma'} , \eqa describing staggered
magnetization in the flux phase of the Heisenberg
model\cite{Nagaosa_Flux}. Because the flux phase results in
massless Dirac fermions near the nodal points, the $QED_3$ with
$S\chi{S}B$, ${\cal L}_{QED} =
\bar{\psi}_{\sigma}\gamma_{\mu}(\partial_{\mu} -
ia_{\mu})\psi_{\sigma} + m_{\psi}\bar{\psi}_{\sigma}\psi_{\sigma}
+ \frac{1}{2g^2}|\partial\times{a}|^2$ can be derived from Eq.
(E1). We perform bosonization by attaching a flux to the fermionic
spinon, $f_{i\sigma} =
b_{i\sigma}exp\Bigl[i\sum_{j}\Theta(i-j)(n_{j\uparrow} +
n_{j\downarrow})\Bigr]$, where $b_{i\sigma}$ is a statistically
transmuted bosonic spinon with $n_{i\sigma} =
f^{\dagger}_{i\sigma}f_{i\sigma} =
b^{\dagger}_{i\sigma}b_{i\sigma}$\cite{CS_Boson}. Our strategy is
to rewrite Eq. (E1) in terms of the newly introduced boson fields
$b_{i\sigma}$, \bqa && H_{eff} = -
\chi_{0}\sum_{<i,j>}(b_{i\sigma}^{\dagger}e^{ia_{ij}}e^{ia_{ij}^{cs}}b_{j\sigma}
+ h.c.) \nn && - m\sum_{<i,j>}(-1)^{i}b_{i\sigma}^{\dagger}{\vec
\tau}^{3}_{\sigma\sigma'}b_{i\sigma'} \eqa with the constraint of
$2\theta{b}_{i\sigma}^{\dagger}b_{i\sigma}{\hat z} = ({\vec
\nabla}\times{\vec a}^{cs})_{i}$, arising from the flux
attachment. Here $a_{ij}^{cs}$ is a Chern-Simons gauge field with
a statistical angle $\theta = \pi$, guaranteeing the statistical
transmutation of the fermionic spinons. Eq. (E2) can be written to
be in a general gauge in the continuum limit \bqa && {\cal L} =
|(\partial_{\mu} - ia_{\mu} - i{a}^{cs}_{\mu})\phi_{\sigma}|^{2} +
m_{\phi}^{2}|\phi_{\sigma}|^{2} +
\frac{u_{\phi}}{2}|\phi_{\sigma}|^{4} \nn && +
\frac{1}{2g^2}|\partial\times{a}|^{2} +
\frac{i}{4\theta}\epsilon_{\mu\nu\lambda}a^{cs}_{\mu}\partial_{\nu}a^{cs}_{\lambda}
, \eqa where the boson field $b_{\sigma}$ is replaced with a
coarse grained field $\phi_{\sigma}$ in the continuum limit.
Comparing Eq. (E3) with the $QED_3$ with $S\chi{S}B$, the Dirac
Lagrangian in the $QED_3$ is replaced with the Klein-Gordon one in
the same dispersion relation {\it owing to the bosonic statistics
of the $\phi_{\sigma}$ field in the flux phase}. $m_{\phi}$ is a
phenomenological boson mass associated with the magnetization $m$
in Eq. (E1) and $u_{\phi}$, strength of local interactions.
$m_{\phi}$ is assumed to be a function of the fermion flavor
number $N$. As $N$ approaches the critical value $N_c$, $m_{\phi}$
goes to zero. The above bosonization procedure can be found in
Ref. \cite{Fisher}, but in an opposite direction (boson to fermion
instead of fermion to boson). The main point is that {\it at the
$QCP$ ($m_{\phi} \rightarrow 0$) the Chern-Simons gauge field does
not play any roles}\cite{Fisher}. Shifting the gauge field
$a_{\mu}$ to $c_{\mu} = a_{\mu} + a^{cs}_{\mu}$ and integrating
over the Chern-Simons gauge field, we obtain a higher order
derivative of the gauge field $c_{\mu}$,
$(\partial\times{c})\cdot(\partial\times\partial\times{c})$. This
higher order derivative term is irrelevant at the $QCP$ in the
$RG$ sense. The resulting effective field theory is obtained to be
${\cal L}_{QCP} = |(\partial_{\mu} - ic_{\mu})\phi_{\sigma}|^{2} +
m_{\phi}^{2}|\phi_{\sigma}|^{2} +
\frac{u_{\phi}}{2}|\phi_{\sigma}|^{4} +
\frac{1}{2g^2}|\partial\times{c}|^{2}$. Remarkably, this critical
field theory has the same form as that derived in a different way
earlier\cite{LQCP}. It is important to realize that this effective
Lagrangian can be {\it physically} meaningful {\it only} at the
$QCP$ because the compactness of the gauge field $c_{\mu}$ can be
irrelevant {\it only} at the $QCP$. Remember that away from the
$QCP$ this effective field theory becomes unstable in the $RG$
sense owing to spinon condensation in the case of $m_{\phi}^{2} <
0$ and instanton condensation in the case of $m_{\phi}^{2} > 0$,
respectively. Especially, the case of $m_{\phi}^{2} > 0$
corresponds to the $ASL$ in the fermion representation. This
discussion shows that the statistics of spinons is fermionic in
the $ASL$.


\begin{thebibliography}{9}
\bibitem{Hubbard_Sigma} K. Borejsza and N. Dupuis, Phys. Rev. B
{\bf 69}, 085119 (2004), and references therein; In this paper the
internal U(1) gauge field $a_{\mu}$ is not introduced since the
U(1) slave boson representation is not utilized.
\bibitem{CS_Boson} This transformation can be justified by the
single occupancy constraint in the Heisenberg model derived from
the Hubbard model in the limit of strong repulsive interactions.
\bibitem{LQCP} If the rigid rotor constraint $|z_{\sigma}|^{2} =
1$ is a little bit relaxed in the critical field theory ${\cal
L}_{QCP} = \frac{Nm_{\psi}^{c}}{2\pi} |(\partial_{\mu} -
ic_{\mu})z_{\sigma}|^{2} + \frac{1}{2g^2}|\partial\times{c}|^2$,
the critical field theory can be written as ${\cal L}_{QCP} =
|(\partial_{\mu} - ic_{\mu})z_{\sigma}|^{2} +
m_{z}^{2}|z_{\sigma}|^{2} + \frac{u_{z}}{2}|z_{\sigma}|^{4} +
\frac{1}{2g^2}|\partial\times{c}|^2$, where $m_{z}^{2} \sim
m_{\psi}^{c} - m_{\psi}$ is an effective spinon mass and $u_{z}$,
strength of local interactions\cite{Senthil_deconfinement}.
\end{thebibliography}
\end{document}